\documentclass[aip,jcp,preprint,superscriptaddress]{revtex4-1}
\usepackage{amsmath}    
\usepackage{graphicx}   
\usepackage{color}      
\usepackage{subfig}     
\usepackage{units,nicefrac}
\usepackage{hyperref}

\begin{document}

\title{Electronic excitations of potassium intercalated manganese phthalocyanine investigated by electron energy-loss spectroscopy}
\author{Benjamin Mahns}
 \email{b.mahns@ifw-dresden.de}
\affiliation{IFW Dresden, P.O. Box 270116, D-01171 Dresden, Germany}
\author{Friedrich Roth}
\affiliation{IFW Dresden, P.O. Box 270116, D-01171 Dresden, Germany}
\author{Mandy Grobosch}
\affiliation{IFW Dresden, P.O. Box 270116, D-01171 Dresden, Germany}
\author{Dietrich R.T. Zahn}
\affiliation{Institute of Physics, Chemnitz University of Technology, Reichenhainer Str. 70, D-09107 Chemnitz, Germany}
\author{Martin Knupfer}
\affiliation{IFW Dresden, P.O. Box 270116, D-01171 Dresden, Germany}
\date{\today}

\date{\today}

\begin{abstract}
The electronic excitations of manganese phthalocyanine (MnPc) films were studied as a function of potassium doping using electron energy-loss spectroscopy in transmission. Our data reveal doping induced changes in the excitation spectrum, and they provide evidence for the existence of three doped phases: K$_1$MnPc, K$_2$MnPc, and K$_4$MnPc. Furthermore, the addition of electrons first leads to a filling of orbitals with strong Mn $3d$ character, a situation which also affects the magnetic moment of the molecule.
\end{abstract}

\maketitle

\section{Introduction}

Molecular thin films consisting of metal phthalocyanines are well known due to their particular electronic properties, which are both of
fundamental as well as of applied relevance.\cite{McKeown1998,Pope1999,A.W.Snow1989}  They, for instance, are successfully used in organic electronic or optoelectronic devices. In their
pristine form, metal phthalocyanines are semiconductors. Moreover, due to their relatively open crystal structure their properties can be tuned
by the incorporation of electron acceptors and donors, and, for example, electrical conduction can be induced by the incorporation of alkali
metals.\cite{Craciun2006,Craciun2005} In addition, the incorporation of transition metal ions in the center of a phthalocyanine ring results in the formation of small,
well defined molecules with a particular magnetic moment. Thus, transition metal phthalocyanines can be regarded as simple model compounds for
the investigation of the fundamental electronic, and magnetic properties of many other transition metal containing molecules, among them
molecular magnetic complexes or molecular magnets.\cite{Kahn1993}

\par

Manganese phthalocyanine (MnPc) is one of these interesting and fundamental compounds.\cite{Mitra1983,Taguchi2006,Fu2007} For instance, MnPc is
characterized by an unusual S = 3/2 spin state of the Mn$^{2+}$ ion \cite{Mitra1983,Taguchi2006,Barraclough1970}, and MnPc has been referred to
as a typical example of a molecular magnet.\cite{Kahn1993} Also, in MnPc the Mn center has a formal 3$d^5$ electronic configuration, and the Mn
3$d$ orbitals are expected to lie close to the chemical potential.\cite{Reynolds1991,Liao2005,Calzolari2007} Recently, we demonstrated that the
occupied electronic structure as well as the electronic excitation spectrum of MnPc is clearly different from all other transition metal
phthalocyanines, which most likely is a direct consequence of the Mn 3$d$ orbitals.\cite{Kraus2009,Grobosch2010,GroboschCPL} In this context, it
is also interesting to investigate the changes that are induced in MnPc by doping with e.\,g. alkali metals.

\par

The physical properties of such doped molecular materials are closely related to the fact whether particular doped phases exist. The detailed microscopic understanding of the physical behavior of the material in question requires the knowledge of the underlying structure and stoichiometry. Recently, it was demonstrated that upon potassium doping of ZnPc, CuPc, and FePc two phases with K$_2$MPc and K$_4$MPc (M =
Zn, Cu, Fe) stoichiometries are formed.\cite{Giovanelli2007,Flatz2007,Roth2008} This parallels the situation as it was found for other potassium doped molecular materials,
e.\,g. alkali metal doped fullerenes.\cite{Weaver1993,Knupfer1994,Poirier1995,Gunnarsson1997} In addition, potassium doping of CoPc leads to a change of the Co spin and charge in the center of
the molecule since the first electron transferred to a CoPc molecule will occupy a Co 3$d$ orbital \cite{Aristov2011}, i.\,e. charge transfer also might be used to
access the magnetic degrees of freedom of such molecules.

\par

In this contribution, we report on an investigation of the electronic properties of potassium doped MnPc films using electron energy-loss
spectroscopy (EELS) in transmission. We discuss the changes that are induced in the electronic excitation spectrum as a function of doping, and
we provide evidence for the formation of three doped phases with K$_1$MnPc, K$_2$MnPc, and K$_4$MnPc composition. Moreover, our data also
indicate filling of a metal 3$d$ orbital upon doping, which then must be accompanied by a change of the magnetic moment of the MnPc molecule in
the charged state.


\section{Experimental}

For our investigations, thin free-standing films of MnPc were prepared under ultra high vacuum conditions (UHV) by thermal evaporation onto
a KBr (100) substrate kept at room temperature. The deposition rate was about $\unit[0.5]{nm/min}$ as measured using a quartz microbalance, and
the resulting film thickness was about 100\,nm. Subsequently, the films were floated off in distilled water, mounted onto standard electron
microscopy grids, and transferred into a purpose built spectrometer  for EELS \cite{Fink1989}, where \textit{in situ} also the doping procedure
was carried out. Prior to doping, the MnPc films were characterized using electron diffraction which showed that the films are essentially
polycrystalline but well textured and mainly consist of the so called $\alpha$-polymorph.\cite{Gould1996,Hoshino2003} Potassium was added from a
commercial SAES source (SAES GETTERS S.p.A., Italy). The doping took place at room temperature and under UHV conditions in several steps by
evaporation of potassium from the getter source at a current of $\unit[6]{A}$ and a source-sample distance of about $\unit[30]{mm}$. We note that this small distance might induce slight heating of the sample during potassium doping.

\par

All EELS measurements were carried out in transmission and at room temperature using a dedicated $\unit[170]{keV}$ spectrometer described
elsewhere.\cite{Fink1989} We note that at this high primary beam energy only singlet excitations are possible. EELS measures the so called loss
function, Im $(-1/\epsilon(\textbf{q},\omega))$ where $\epsilon(\textbf{q},\omega)$ is the dielectric function. It thus contains information on
the dielectric response of the material under investigation and it can also probe the excitations as a function of momentum transfer \textbf{q}
even far away from the optical limit.\cite{Knupfer2000,Kramberger2008,Schuster2007} The loss function of our samples was measured for a
momentum transfer of \textbf{q}=$\unit[0.1]{\mathring{A}^{-1}}$, parallel to the film surface. In order to obtain a direction independent core
level excitation information, we have determined the core level data for three different momentum directions such that the sum of these spectra
represent a polycrystalline average also for the textured samples.\cite{Egerton1986} The energy and momentum resolutions were
$\unit[90]{meV}$ and $\unit[0.03]{\mathring{A}^{-1}}$ for the loss function and diffraction measurements and $\unit[200]{meV}$ and
$\unit[0.03]{\mathring{A}^{-1}}$ in the case of the measured core level excitations. Furthermore, as molecular crystals often are damaged by
fast electrons, we repeatedly checked our samples for any sign of degradation and did not consider those with changes in electronic or
diffraction spectra. For further details of the sample preparation and the experimental technique we refer the reader to previous publications.\cite{Fink1989,Knupfer2001}


\section{Results and discussion}

 \begin{figure}[h]
 \centering
  \subfloat{\includegraphics[width=0.5\textwidth]{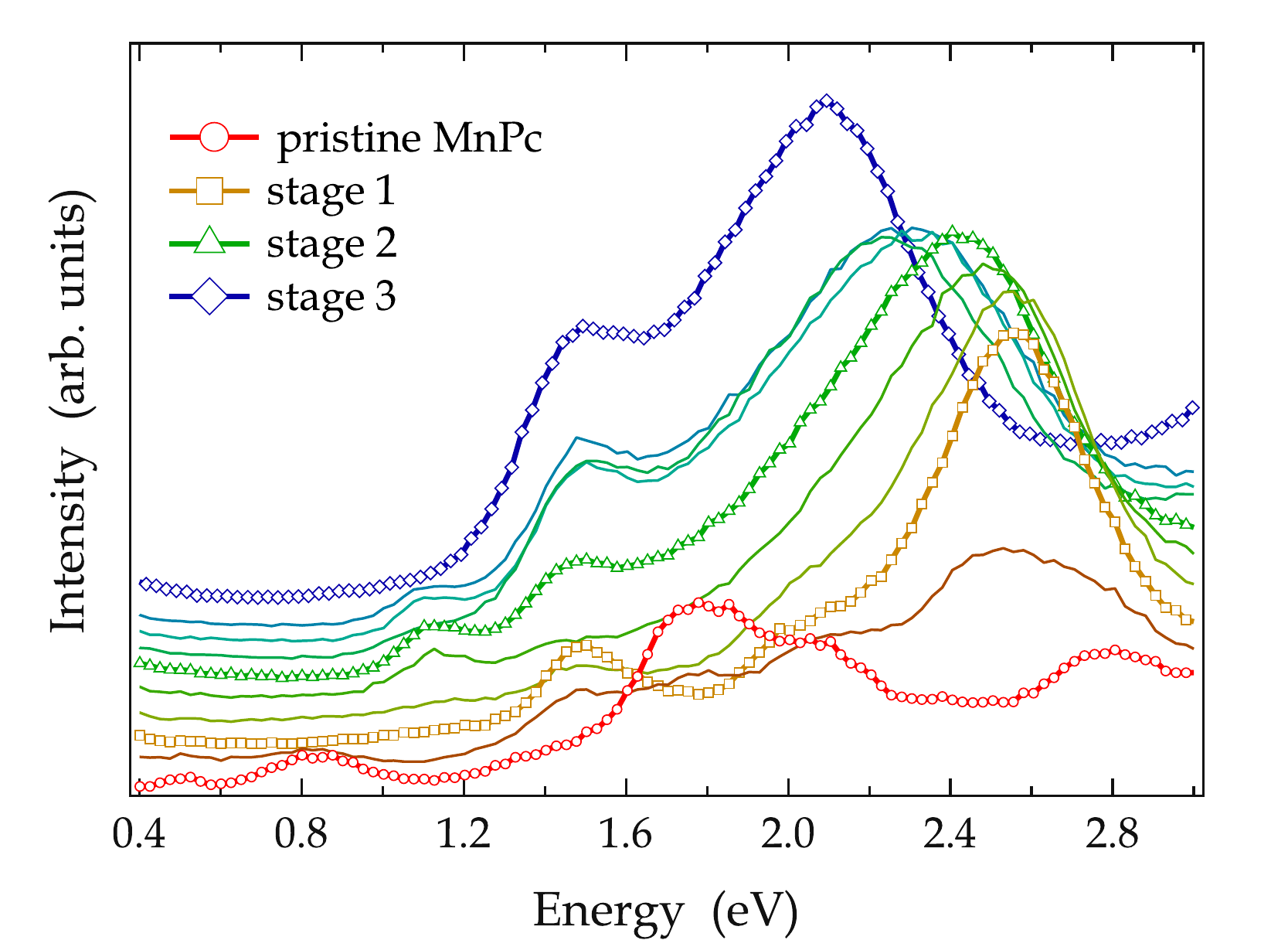}}
  \subfloat{\includegraphics[width=0.5\textwidth]{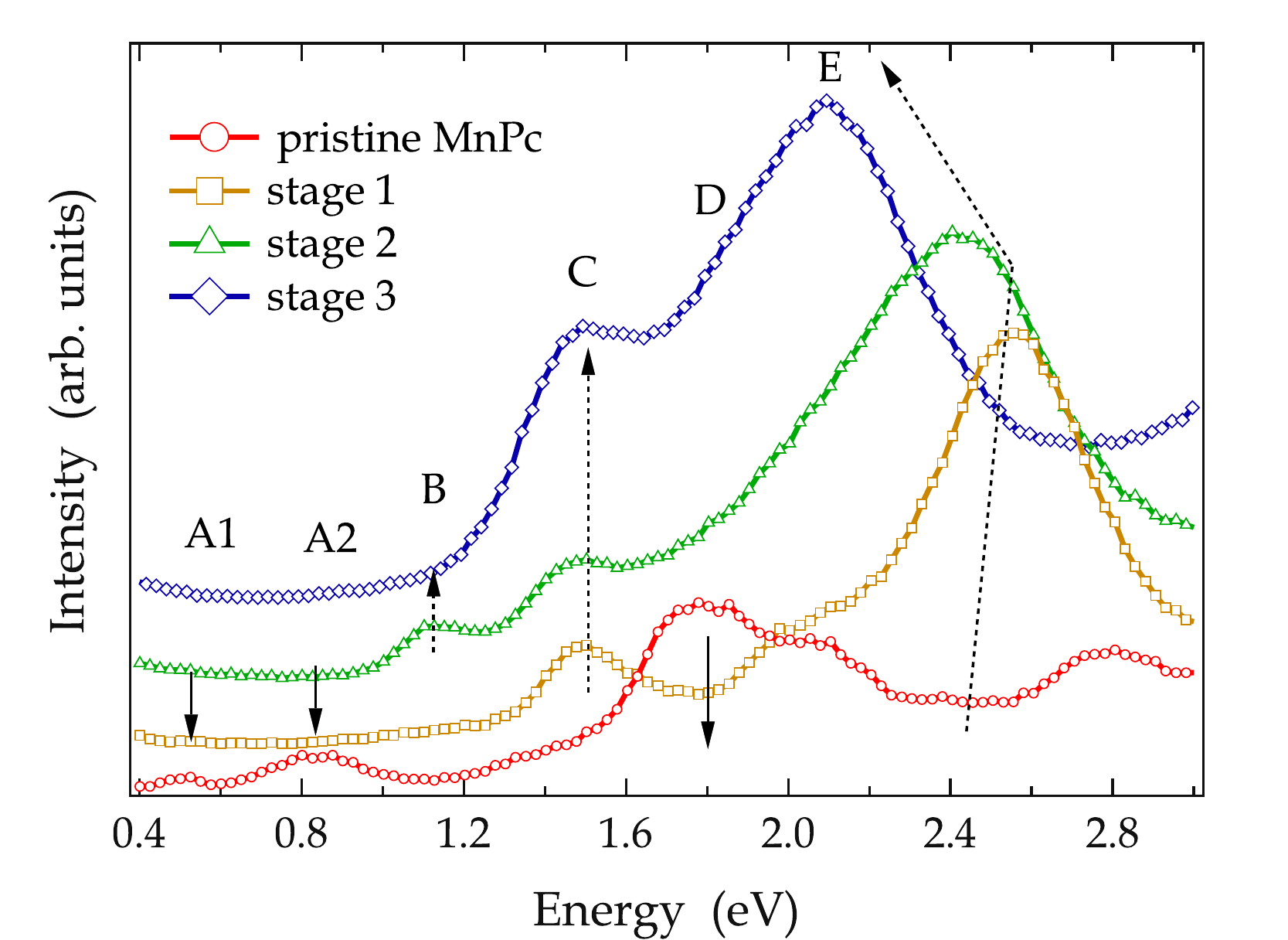}}
 
\caption[]{Evolution of the loss function of MnPc upon potassium doping. The potassium content increases from bottom to top. In the left panel, thicker lines with
 symbols indicate particular doping stages, where the spectra indicate the formation of K doped MnPc phases (see text). In the right panel, we only show the spectra of these K doped MnPc phases. The
 various excitation features are labelled with capital letters (see discussion for further details).} \label{fig1}
 \end{figure}

We start the presentation of our results with the evolution of the electronic excitation spectra of MnPc upon potassium doping, which is shown
in Fig.\,\ref{fig1}. These data were taken with a momentum transfer of 0.1 \AA$^{-1}$ and we note that for such a small momentum transfer EELS probes
dipole allowed excitations and the data are thus equivalent to those from optical studies.\cite{Fink1989} The loss function of pristine MnPc
(bottom curve in Fig.\,\ref{fig1}) consists of several excitation features at 0.5, 0.8, 1.4, 1.8, 2.1, and 2.8\,eV, which were already reported
earlier.\cite{Kraus2009} The electronic excitation features at 1.8 - 2\,eV are associated with the well known Q band common to many
phthalocyanines.\cite{McKeown1998,Pope1999,A.W.Snow1989} This Q band in the solid state is broad and structured because of vibronic satellites
and the impact of solid state effects.\cite{A.W.Snow1989,Peisert2002a} It has its origin in excitations from the occupied molecular orbital with
a$_{1u}$ symmetry to an unoccupied molecular orbital with e$_{g}$ symmetry.\cite{McKeown1998,Pope1999,A.W.Snow1989,Liao2005,Kraus2009} The exact
origin of the excitations at lower energies is unclear to date. They have been discussed in terms of the contribution of Mn 3$d$ states to the
molecular orbitals close to the chemical potential in MnPc \cite{Kraus2009} but there is no full, consistent understanding of what can be
observed in Fig.\,\ref{fig1} for pristine MnPc.

\par

Upon potassium addition we observe significant changes in the excitation spectra. First, the two features A1 and A2 and the Q-band excitation
(D) disappear while new spectral structures show up at 1.5\,eV (C) and about 2.6\,eV (E). Further doping leads to an energetic down-shift of
feature E and the appearance of a further excitation at about 1.1\,eV (B). The intensity of feature B reaches a maximum at intermediate doping
and then decreases in intensity, while feature C grows slightly and feature E continues to shift to lower energies.

\par

The data in Fig.\,\ref{fig1} indicate the existence of distinct K-MnPc compositions (stages 1, 2 and 3) with particular stoichiometry and optical
properties. Stage 1 is reached when the low energy structures A1 and A2 have vanished and features C and E have developed at 1.5 and 2.6\,eV,
respectively. The addition of more potassium induces the formation of stage 2, which is signalled by the appearance of feature B and the
downshift of feature E. Finally, in stage 3 feature B disappears again, feature C increases in intensity and feature E reaches its final energy
position at about 2.1\,eV. In general, the formation of distinct potassium doped MnPc phases parallels the situation in other transition
phthalocyanines upon addition of potassium, where it was reported that phases with compositions K$_2$ZnPc, K$_4$ZnPc, K$_2$CuPc, K$_4$CuPc,
K$_2$FePc, and K$_4$FePc exist.\cite{Giovanelli2007,Flatz2007,Roth2008} However, we emphasize that the data in Fig.\,\ref{fig1} indicate three potassium
doped MnPc structures, one more than for the other phthalocyanines.

\par

 \begin{figure}[h]
  \centering
  \subfloat{\includegraphics[width=0.5\textwidth]{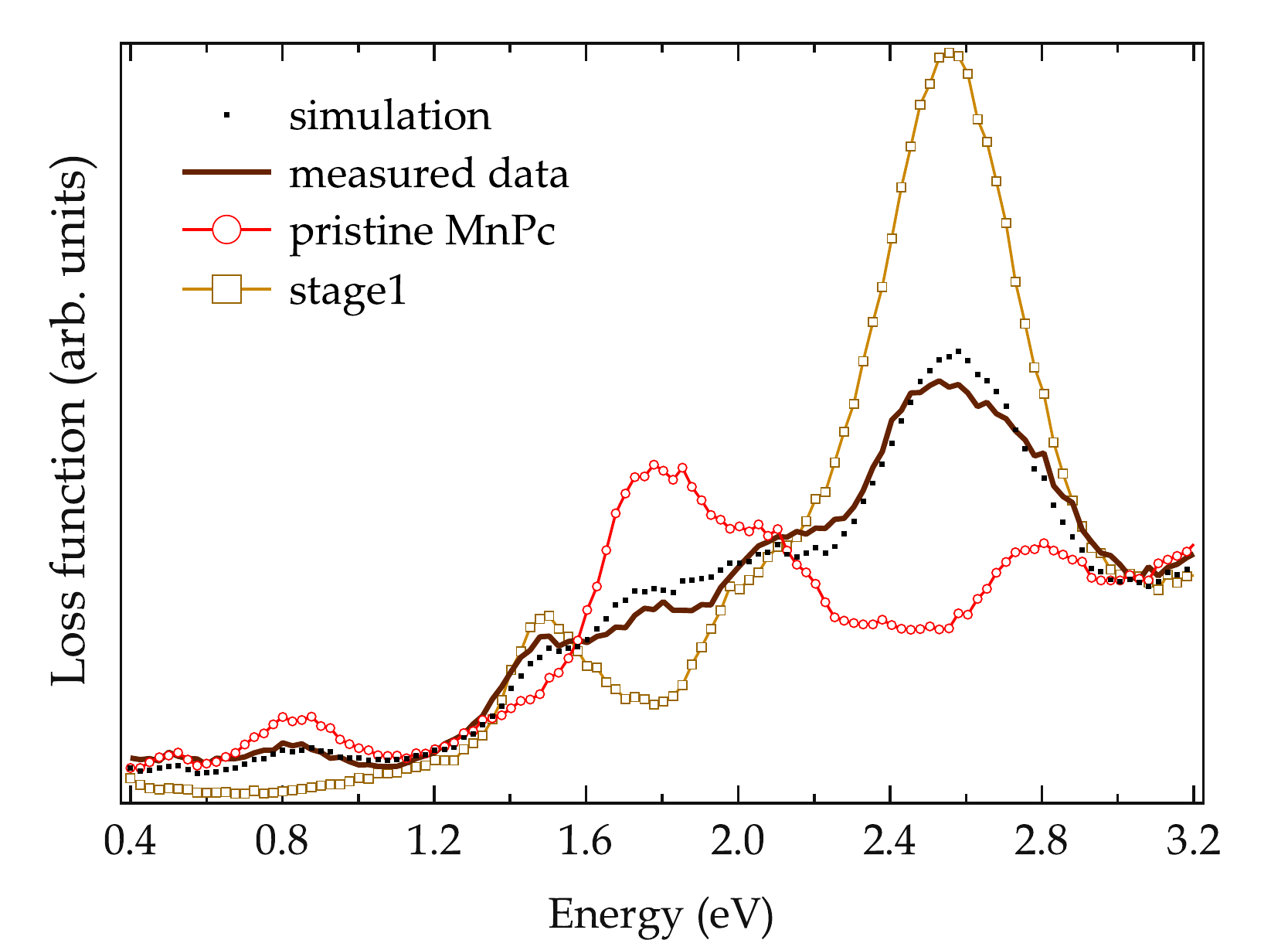}}
  \subfloat{\includegraphics[width=0.5\textwidth]{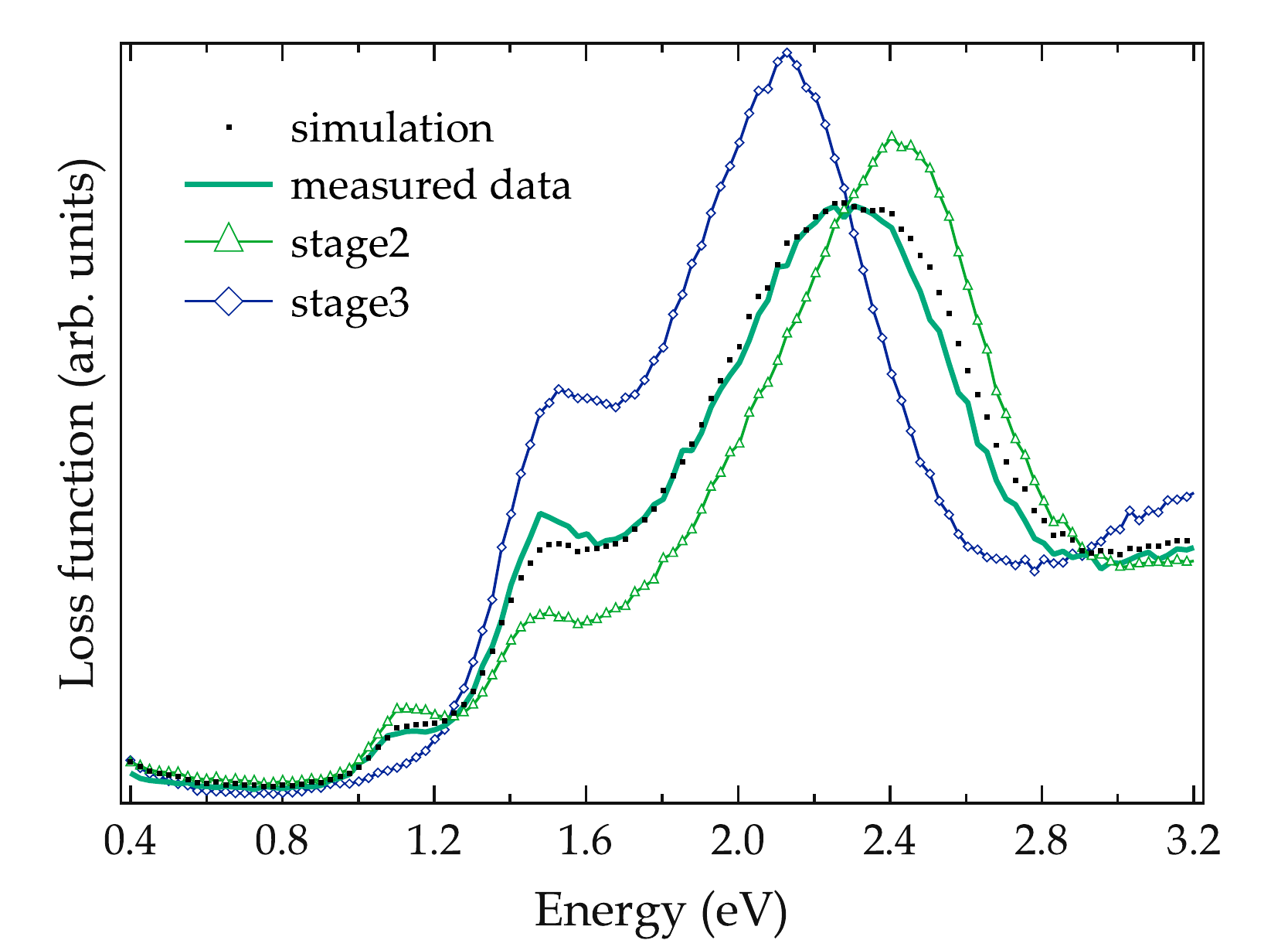}}

 \caption[]{Comparison of the loss function of the observed K doped MnPc stages (see text) and data for samples with potassium content
 between pristine MnPc and stage 1 and between stage 2 and stage 3, respectively. The data for these intermediate compositions
have been modelled by a weighted superposition of the corresponding intensities of neighboring stages. In the left panel the weights are 0.5 for
both stages, while in the right panel they were 0.65 (stage 2) and 0.35 (stage 3).}
 \label{fig2}
 \end{figure}

Our conclusion that stages 1, 2 and 3 as described above correspond to stable, potassium doped phases of MnPc requires that the spectra
in-between can be described by a superposition of the spectra of the corresponding phases in the direct neighborhood, in analogy to what was observed for K doped CuPc. In Fig.\,\ref{fig2} we show a comparison of corresponding data for potassium contents between undoped
MnPc and stage 1 as well as stage 2 and 3, respectively, with modelled data obtained by a weighted sum of the data for the respective phases.
The agreement of measured and modelled data is very good, which corroborates our conclusion that these phases are formed.

\par

 \begin{figure}[ht]
  \centering
  \subfloat{\includegraphics[width=0.5\textwidth]{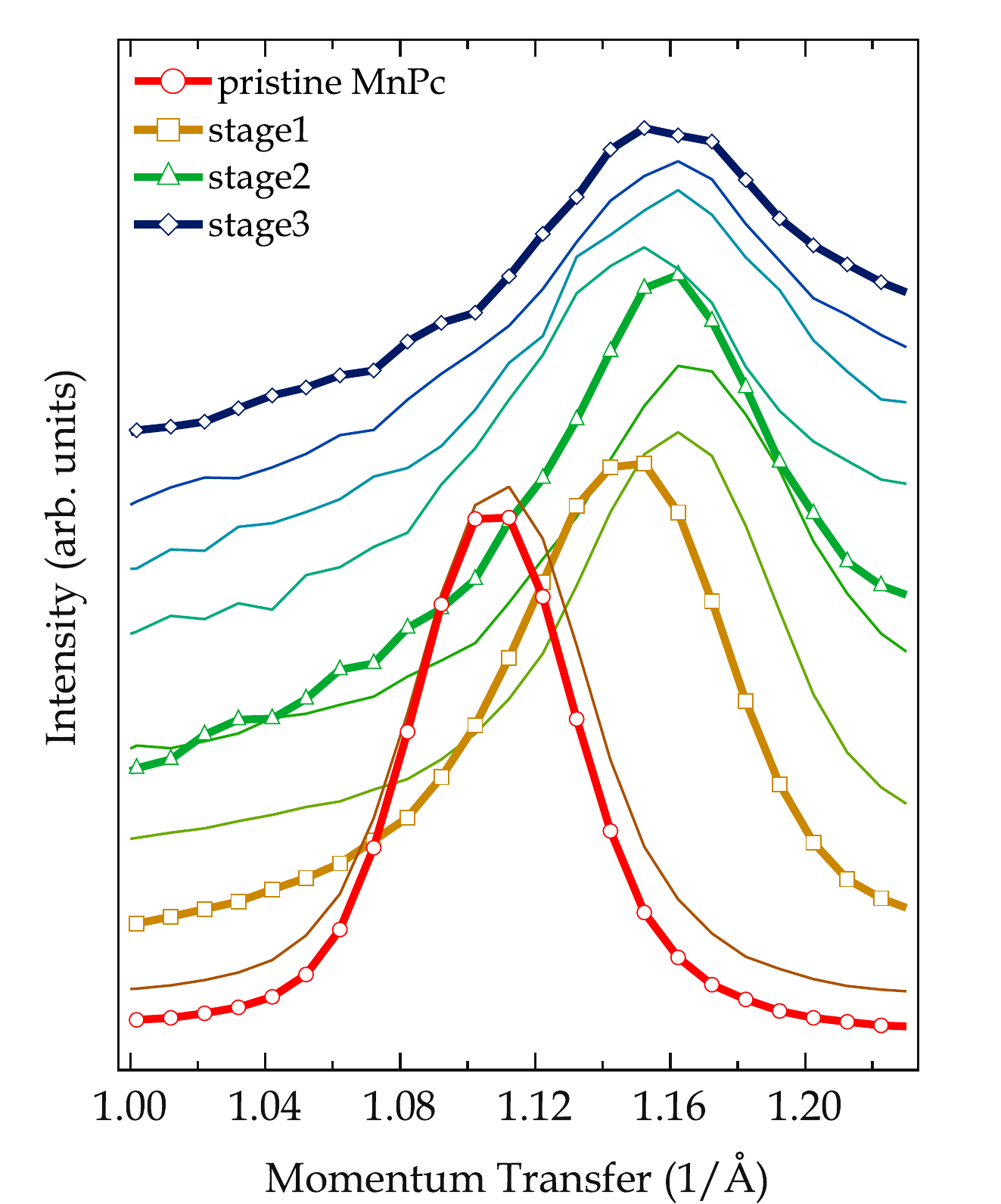}}
  \subfloat{\includegraphics[width=0.5\textwidth]{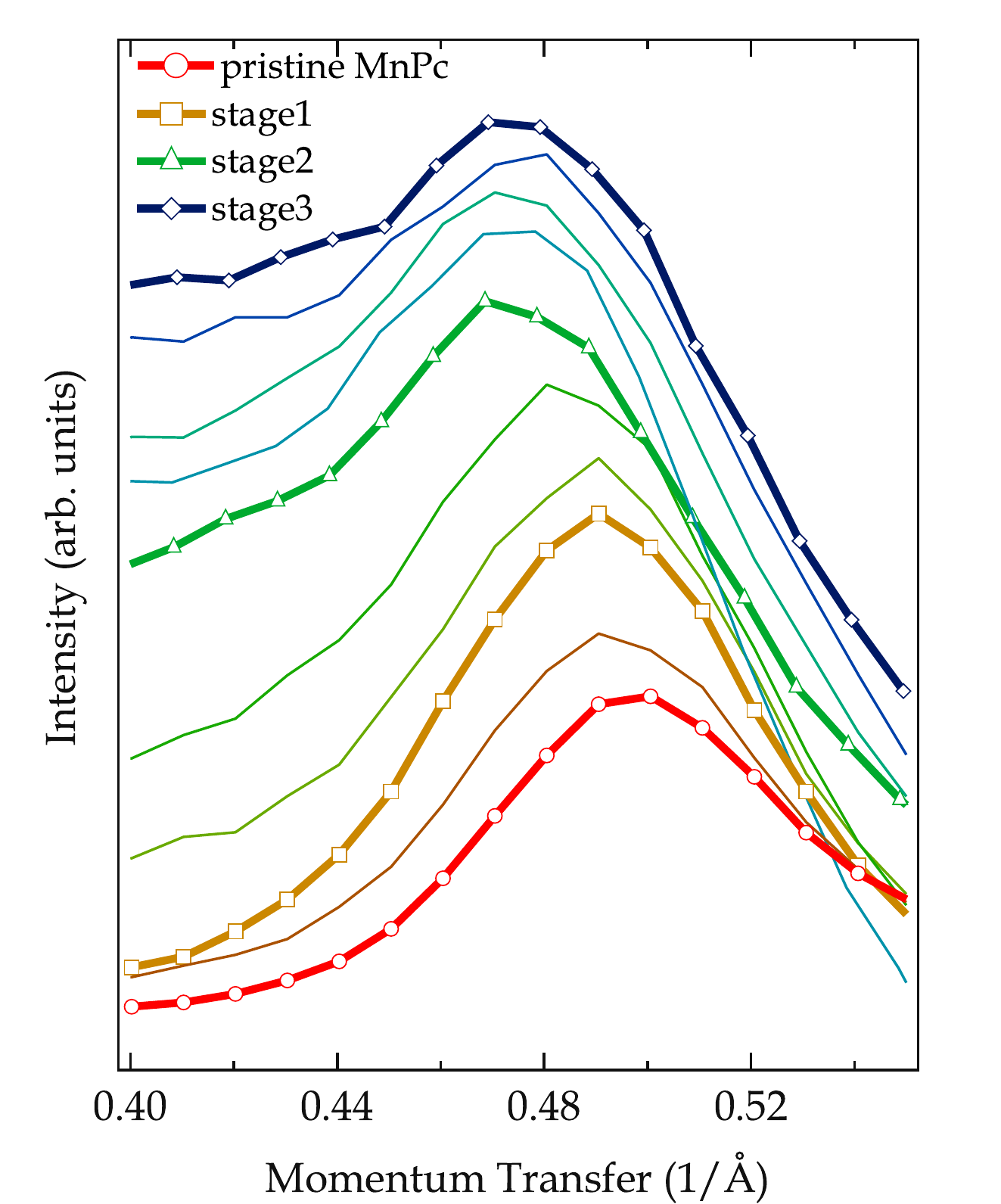}}

 \caption[]{Doping dependence of the electron diffraction profiles around 1.1 and 0.48\,\AA$^{-1}$.} \label{fig3}
 \end{figure}

Moreover, evidence that stable phases are formed can also be read off the evolution of the electron diffraction profiles of our samples. In Fig.\,\ref{fig3}, we show the doping induced changes around 1.1\,\AA$^{-1}$ and 0.5\,\AA$^{-1}$. The diffraction maximum at 1.12\,\AA$^{-1}$ (left panel of Fig.\,\ref{fig3}) in the undoped MnPc samples is due to the (201) Bragg peak of the $\alpha$-polymorph of MnPc \cite{Gould1996,Hoshino2003}, and it vanishes
upon potassium addition. Reaching doping stage 1, a new diffraction peak at 1.15 \AA$^{-1}$ is observed, which is further shifted going to the
doped stages 2 and 3. The diffraction peak at 0.5\,\AA$^{-1}$ in pristine MnPc (right panel of Fig.\,\ref{fig3}) corresponds to the (100) Bragg peak and
shifts to lower values upon doping. For stage 1 it is found at about 0.48\,\AA$^{-1}$, while for the two other stages it is at about 0.47\,\AA$^{-1}$.

 \begin{figure}[ht]
 \centering
 \includegraphics[width=0.5\textwidth]{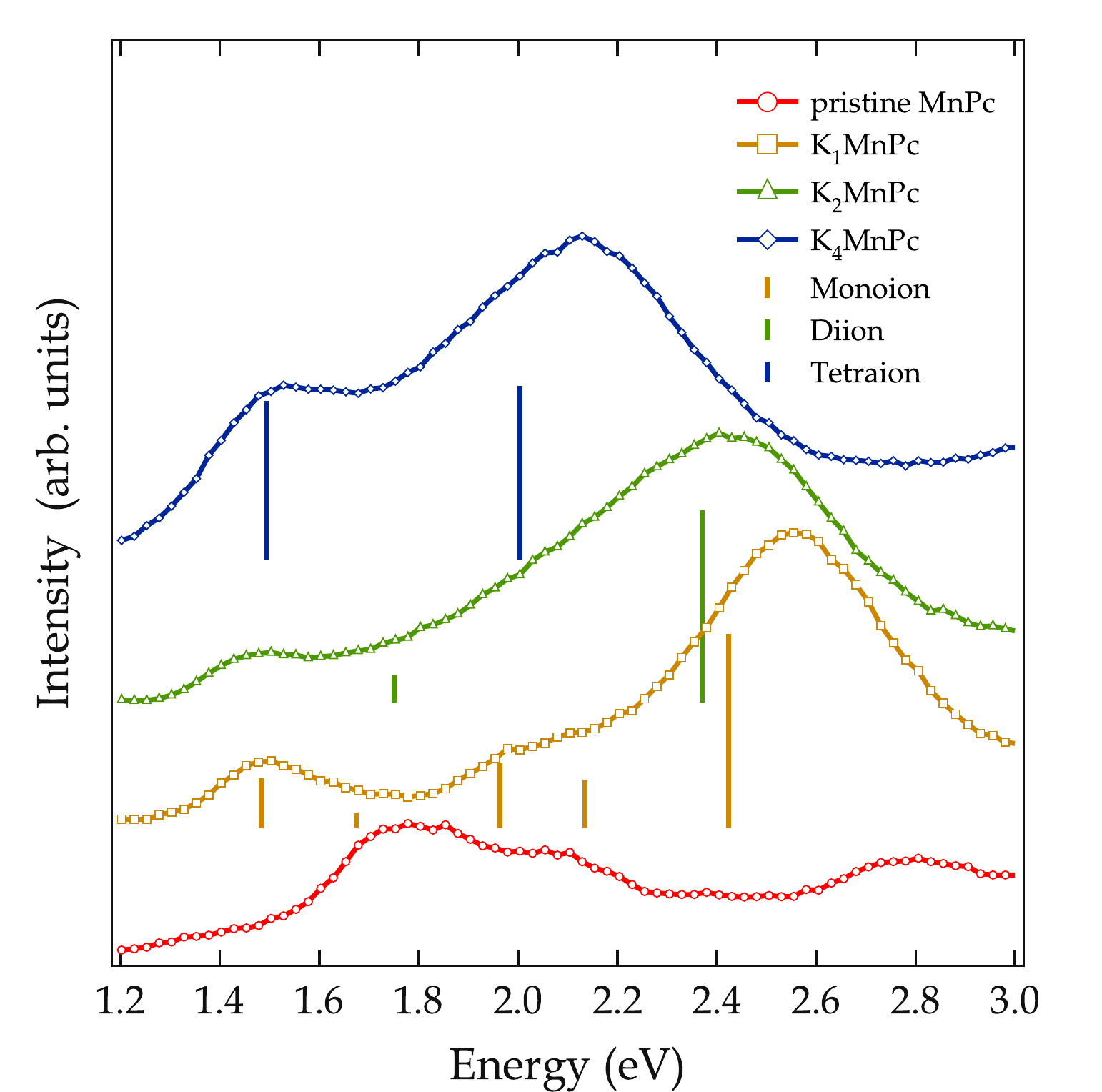}
 \caption{Comparison of the loss function of K$_x$MnPc phases with $x=0,1,2,4$. The bars indicate the excitation energies as seen in optical
 absorption for electrochemically reduced MnPc$^-$, MnPc$^{2-}$, and MnPc$^{4-}$ molecules in solution.\cite{Clack1972,Minor1985}}
 \label{fig4}
 \end{figure}

Further support for our conclusion, that stages 1, 2, and 3 are formed, can be obtained by a comparison of our data to the optical absorption
taken for electrochemically charged MnPc molecules in solution.\cite{Clack1972,Minor1985} This is depicted in Fig.\,\ref{fig4}, which summarizes the excitation spectra obtained for
the undoped MnPc and the three potassium doped stages and compares those to the excitation energies obtained for the MnPc$^-$ monoion,
MnPc$^{2-}$ diion and MnPc$^{4-}$ tetraion in solution.\cite{Clack1972,Minor1985} In consideration of the broadening upon going from single molecules in solution
to the solid state, and a possible different excitation energy due to different screening effects, the data presented in Fig.\,\ref{fig4} indicate the
same trend for potassium doping in the solid and negative charging in solution. We therefore assign our doped phases to K$_1$MnPc (stage 1), K$_2$MnPc (stage 2) and K$_4$MnPc (stage 3).

\par

\begin{figure}[ht]
 \centering
 \includegraphics[width=0.7\textwidth]{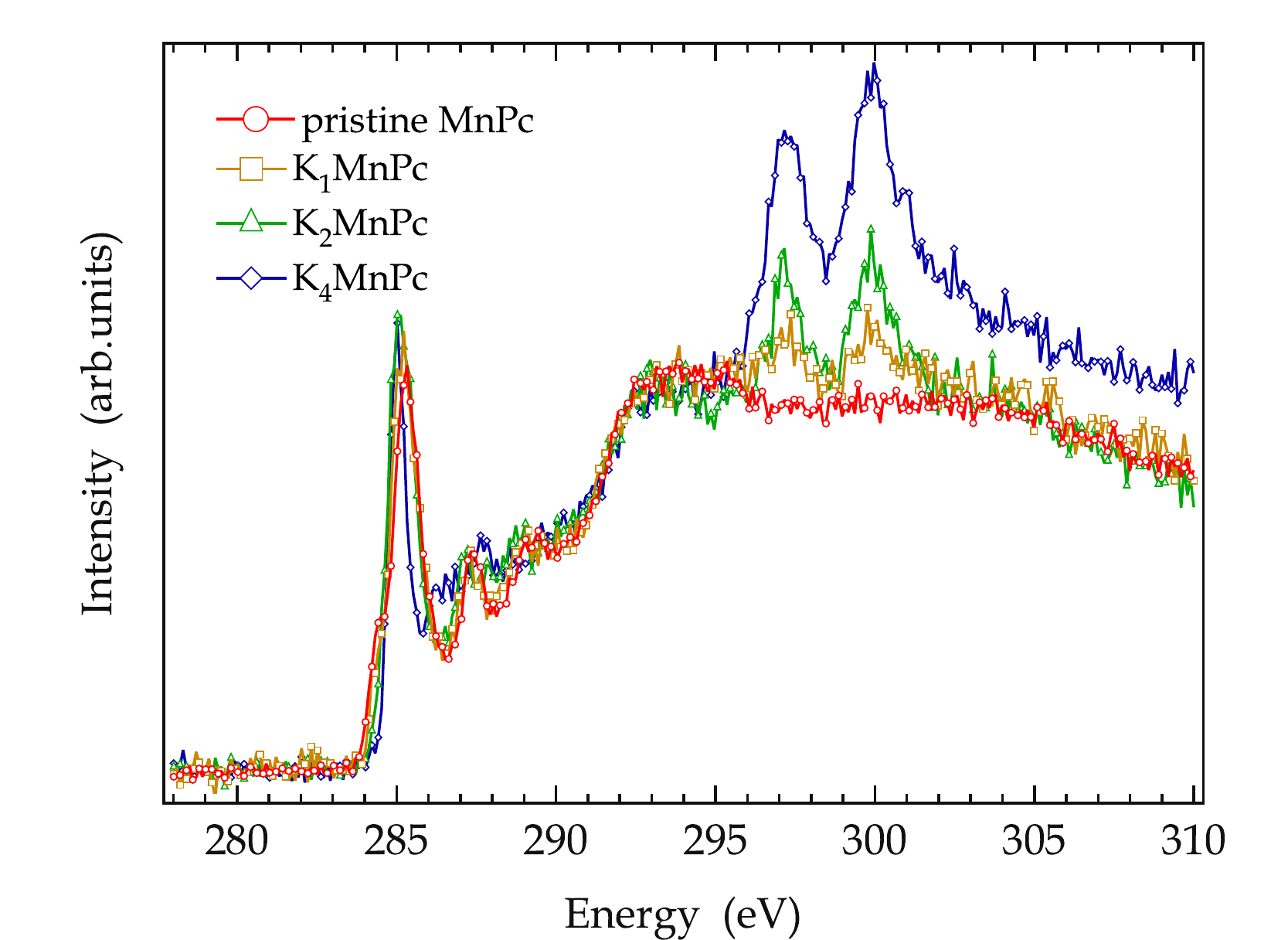}
 \caption[]{C 1$s$ and K2 $p$ core level excitations of pristine MnPc and the potassium doped MnPc phases.} \label{fig5}
 \end{figure}

Finally, the amount of potassium in the samples can be determined by EELS core level excitation measurements at the C 1$s$ and K 2$p$ core levels. In Fig.\,\ref{fig5} we show these core excitations for undoped MnPc and the three identified doped phases. In general, the data in Fig.\,\ref{fig5} arise from
excitations of carbon 1$s$ electrons into unoccupied $\pi^*$ states at 285 - 291\,eV, those from the C 1$s$ level into carbon derived $\sigma^*$
states with the so-called $\sigma$ edge at 291 - 292\,eV, and K 2$p$ to K 3$d$ excitations with maxima at 297.1 and 298.8\,eV.\cite{Flatz2007,Roth2008,Knupfer2001} The data are normalized to the intensity of the first C 1$s$ $\sigma^*$ feature at about 293\,eV which
represents the amount of carbon in the sample. Fig.\,\ref{fig5} clearly shows the increasing potassium content in our samples. An analysis of the
intensities of the K 2$p$ excitation structures in comparison to well characterized potassium doped fullerenes or highly oriented pyrolytic
graphite (HOPG) yields a doping level K$_x$MnPc of $x \sim 0.9, 2.1$ and 4 for our samples. Details of this procedure can be found in previous
publications.\cite{Flatz2007,Roth2008,Knupfer2001,Pichler1999,Liu2003,Liu2004} We estimate the error of this analysis to be $\Delta x$ $\sim$
$\pm 0.2$ (K$_x$MnPc). Thus, this core level analysis corroborates our assignment of the identified potassium doped MnPc phases to K$_1$MnPc,
K$_2$MnPc and K$_{4}$MnPc, respectively.

\par

Consequently, in analogy to other transition metal phthalocyanines also K doping of MnPc results in particular doped phases, but MnPc is the
only phthalocyanine reported so far that also forms a K$_1$MnPc phase. This difference might be closely related to the occupancy and energy
position of the 3$d$ levels of the central transition metal. In the case of MnPc, data from photoemission spectroscopy, EELS and theory clearly
indicate the presence of Mn $3d$ orbitals very close to the chemical potential, and actually at energies in-between the fully occupied $a_{1u}$
and the lowest unoccupied $e_{g}$ ligand molecular orbitals.\cite{Calzolari2007,Kraus2009,Grobosch2010,GroboschCPL} Furthermore, these Mn 3$d$
orbitals are also responsible for a significant change of the ionization potential and most likely also the electron affinity. Previous
electrochemistry investigations have provided evidence that the first oxidation as well as the first reduction state of MnPc molecules have
strong Mn 3$d$ character.\cite{A.W.Snow1989,Liao2005} We therefore infer that the first electron that is transferred by potassium addition to
each MnPc molecule in the solid state will occupy a state with predominant Mn 3$d$ character, which is in agreement with the disappearance of
the low energy electronic excitations as seen in undoped MnPc and which might help to stabilize a phase with corresponding stoichiometry,
K$_1$MnPc. It is interesting to note, that also in potassium doped CoPc films the first electron occupies a metal (Co) 3$d$ state before the
ligand orbitals are filled.\cite{Aristov2011} Further electron transfer caused by further potassium addition then leads to the occupation of the
ligand $e_g$ orbitals with electrons and the formation of the other observed phases. The fact that for all transition metal phthalocyanines
investigated so far the maximal potassium doping is $x$ = 4 is most likely a consequence of steric or geometrical reasons.

\par

Finally, the observed filling of a Mn 3$d$ orbital upon potassium addition demonstrates the formation of Mn(I) centers, which is directly
accompanied by a change of the magnetic moment. This now evinces a way to deliberately change the magnetic properties of such molecules via
charge transfer in thin films. It additionally might have consequences for the charge transport behavior of MnPc films, since electrons injected
into a film might occupy metal 3$d$ rather than ligand $\pi$ orbitals with rather different wave function and hopping probabilities. The
associated variation of the spin state of MnPc can also affect the spin scattering mechanisms in spin-dependent transport.

\section{Summary}

To summarize, we studied the electronic properties of undoped and potassium doped manganese phthalocyanines using electron energy-loss
spectroscopy. Our data allow the identification of three doped phases, K$_1$MnPc, K$_2$MnPc, and K$_{4}$MnPc and show the variations in the
electronic excitation spectra upon doping. In addition, in K$_1$MnPc a MnPc orbital with predominant 3$d$ character is filled, i.\,e. doping can
change the valence and spin state of the Mn center.

\begin{acknowledgments}
We are grateful to R. H\"ubel, S. Leger and R. Sch\"onfelder for technical assistance. Financial support by the Deutsche Forschungsgemeinschaft
within the Forschergruppe FOR 1154 (Project KN393/14) is gratefully acknowledged.

\end{acknowledgments}


%

\end{document}